\documentclass[baaa]{baaa}

\usepackage[pdftex]{hyperref}
\usepackage{subfigure}
\usepackage{natbib}
\usepackage{helvet,soul}
\usepackage[font=small]{caption}

\contriblanguage{0}

\contribtype{3}

\thematicarea{1}

\title{Observando la cromosfera solar en el infrarrojo}

\titlerunning{La cromosfera solar en el IR}

\author{C.~G.~Giménez de Castro\inst{1,2}}
\authorrunning{Giménez de Castro}

\contact{guigue@craam.mackenzie.br}

\institute{
Centro de Rádio Astronomia e Astrofísica Mackenzie, Universidade Presbiteriana Mackenzie, São Paulo, Brasil. \and
Instituto de Astronom\'ia y F\'isica del Espacio, CONICET--UBA, Buenos Aires, Argentina.
}

\resumen{La cromosfera solar ha sido históricamente estudiada a partir
  de líneas espectrales en el visible y el UV, notablemente H$\alpha$,
  Ca~\textsc{ii}, Mg~\textsc{i} y Ly$\alpha$. Recientemente se han
  agregado observaciones espaciales en las longitudes de onda largas
  del UV (304, 1600 y 1700 \AA). Sin embargo, también puede ser
  estudiada en el infrarrojo (IR), tanto en el continuo como en las
  líneas. Estudios en esta banda espectral, que por definición se
  extiende de 1 $\mu$m a 1 mm, son escasos y recientes, habiéndose
  explorado poco sus ventajas. En este trabajo presentamos una
  revisión de lo que se ha hecho y detallamos lo mucho que se puede
  hacer con instrumentos instalados en tierra.  Argentina cuenta con
  un conjunto de telescopios únicos para la observación de la
  cromosfera, algunos con más de 20 años de operación y en proceso de
  renovación, otros recientemente instalados y algunos en
  desarrollo. El panorama es muy alentador y permite prever una fuerte
  cooperación internacional con otros instrumentos en tierra y
  embarcados en sondas espaciales.}

\abstract{The solar chromosphere has historically been studied from
  spectral lines in the visible and UV, notably H$\alpha$,
  Ca~\textsc{ii}, Mg~\textsc{ii} and Ly$\alpha$. Observations at long
  UV wavelengths (304, 1600 and 1700 \AA) from space have been
  recently added.  However, the chromosphere can also be studied in
  the infrared (IR), both in the continuum as in the lines. Studies in
  this spectral band, which by definition extends from 1 $\mu$m to 1
  mm, are scarce and recent, and its advantages having been little
  explored. In this work we present a review of what has been done and
  detail how much can be done with ground-based instruments. Argentina
  has a set of unique telescopes for the observation of the
  chromosphere, some with more than 20 years of operation and in
  process of renovation, others recently installed and still some in
  development. The panorama is very encouraging and allows to
  anticipate a strong international cooperation with other ground and
  space facilities.}


\begin{document}

\maketitle

\section{Introducción}

Conocemos la cromosfera fundamentalmente gracias a observaciones en el
visible de líneas espectrales como H$\alpha$ o C~\textsc{ii}. La
astronomía espacial nos permitió observarla a través de filtros
ultravioleta (UV) en líneas espectrales de Fe~\textsc{i} y
He~\textsc{i}, entre muchas otras, o en el continuo en 1600 y
1700~\AA. Fotosfera y cromosfera pueden ser observadas también en el
infrarrojo (IR), tanto en el contínuo como a través de líneas
espectrales. Sin embargo, a pesar del amplio rango espectral que
abarca el IR, este no ha sido debidamente explorado. Las razones son
múltiples: la tecnología IR era considerada hasta hace pocos años,
militar, por lo tanto cara y de acceso restringido. Observaciones en
estas longitudes de onda \textit{oscuras} requieren de observatorios
en altura, en algunos casos, incluso fuera de la troposfera. Sólo en
los últimos años estas desventajas se han ido revirtiendo.  En este
artículo hacemos una revisión de las observaciones solares en el IR,
destacando las mayores contribuciones. Además describimos la
instrumentación actual y la futura con hincapié en los telescopios
instalados (o a instalar) en Argentina.

\subsection{El Infrarrojo}

Desde el experimento pionero de los hermanos Herschel que los llevó a
especular con la existencia de una radiación \textit{por debajo del
  rojo}, podemos definir al IR como la región espectral invisible con
frecuencias inferiores a las del rojo.  Como toda definición los
límites del IR son arbitrarios y sujetos al área de
actuación. Usando criterios tanto físicos (de los que hablaremos más
abajo) como tecnológicos definimos al rango IR
$$1000 \gtrsim \lambda_\mathrm{IR} \gtrsim 1\ \mu\mathrm{m} \ , \quad
0.3 \lesssim \nu_\mathrm{IR} \lesssim 300\ \mathrm{THz}\ .$$
Este rango de 3 órdenes de magnitud lo dividimos por comodidad en\\
\begin{tabular}{rcc}
          & $\simeq\lambda\ [\mu$m] & $\simeq\nu$~[THz] \\
  \hline
  Próximo (IRP)  &  1 --     5 & 300 -- 60   \\
  Medio (IRM)    &  5 --    30 &  60 -- 10   \\
  Lejano (IRL)   &  30 --  300 &  10 --  1   \\
  submilimétrico (submm) & 300 -- 1000 &   1 --  0,3 \\
\end{tabular}

Sólo recientemente el mercado ha presentado una variedad de oferta
accesible y sin restricciones para este rango de frecuencias. Para el
IRM y el IRL hoy se puede comprar cámaras comerciales a temperatura
ambiente por precios que rondan las decenas de miles de dólares,
también se puede acceder a detectores a los que se les adosa una
electrónica específica incluyendo control de temperatura. En el
submm la tecnología es la de receptores superheterodinos
(\textit{coherentes}) refrigerados o no, aunque todavía no cuenta con
muchos proveedores disponibles. En la última década aparecieron firmas
ofreciendo bolómetros (detectores térmicos \textit{incoherentes}) para
longitudes específicas desde el submm hasta el IRM con gran ancho de
banda (10--20\%~$\times\nu_\circ$).\\

\begin{figure}
\centerline{\includegraphics[width=0.8\textwidth, angle=90]{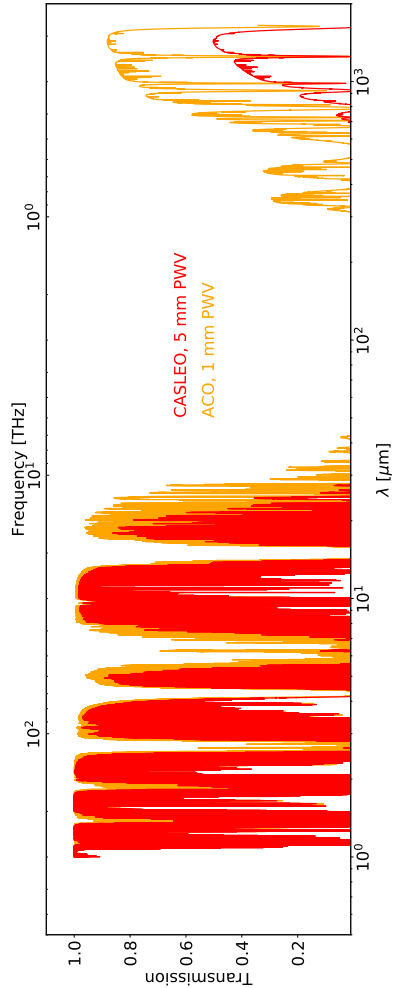}}
\caption{Transmisión atmosférica al cénit en función de la longitud de
  onda para Alto Chorrillos (ACO, Salta) a 4800 m sobre
  el nivel del mar y con un contenido de vapor de agua PWV=1~mm (curva
  naranja) y para el CASLEO a 2500 m sobre el nivel del mar con
  PWV=5~mm (curva roja).}
\label{fig:IRTransmission}
\end{figure}

Otro punto en común del rango de frecuencias que hemos definido como
IR es la absorción atmosférica: la atmósfera terrestre es bastante
opaca a la radiación a lo largo de los tres órdenes de frecuencia. En
la Figura \ref{fig:IRTransmission} mostramos la transmisión
atmosférica al cénit calculada usando el programa
\href{https://atran.arc.nasa.gov/cgi-bin/atran/atran.cgi}{ATRAN} para
dos lugares diferentes: Alto Chorrillos (sitio de instalación del
radioteslecopio Llama) para un contenido de vapor de agua PWV=1~mm y
el CASLEO con PWV=5~mm que son valores típicos de cada lugar. Como se
puede ver, el IRL es completamente invisible y de hecho sólo
observable por encima de la troposfera. El submm, por otra parte, es
razonablemente observable desde ACO entre 1000 y 700~$\mu$m
aproximadamente, después la transmisión cae por debajo del 50\%,
mientras que desde CASLEO la atenuación es bastante fuerte limitando
el flujo mínimo detectable y la calidad de las observaciones sujetas a
las variaciones de la temperatura atmósferica. Ya en el IR Medio y
Próximo tenemos rangos bien definidos y anchos en los que es posible
observar con comodidad desde cualquiera de los dos sitios. La
transmisión \textit{efectiva}, sin embargo, es menor porque hay
millones de líneas de absorción atmosféricas que en la figura se
encuentran superpuestas dificultando su identificación individual. En
este trabajo mos enfocamos en los rangos Medio a submm del IR, es
decir de $\approx 5 \ \mathrm{a} \ 1000\ \mu$m.

\subsection{El continuo}

\begin{figure}
\centerline{\includegraphics[width=0.5\textwidth]{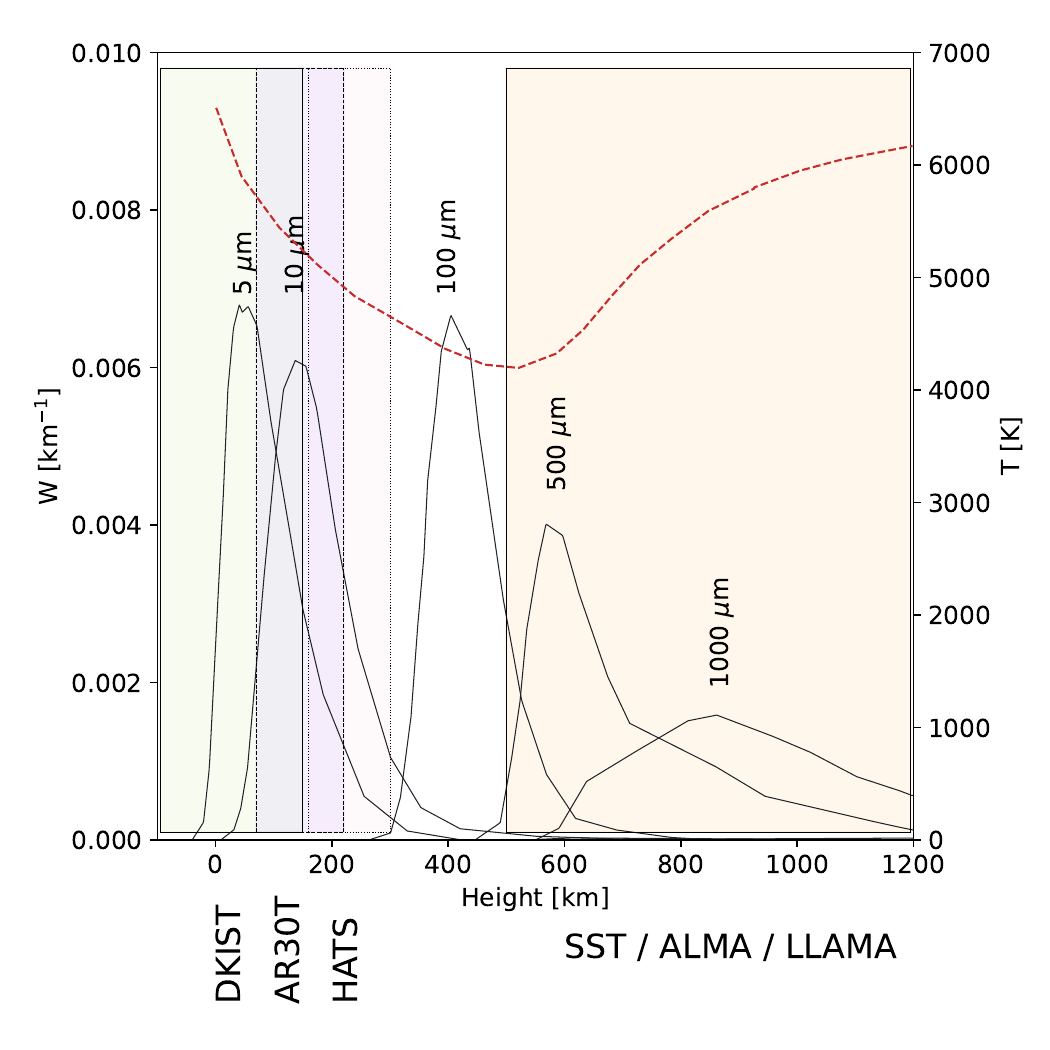}}
\caption{Fuciones de ponderación de la emisión del contínuo para
  diferentes longitudes de onda en función de la altura (curvas
  negras). La curva punteada es la distribución de temperatura del
  modelo VAL-C \citep{VAL81}. Las áreas coloreadas marcan
  aproximadamente el rango de alturas observadas por los diferentes
  instrumentos comentados en este artículo. Fuente: adaptada de
  \cite{Jefferies:1994}.}
\label{fig:valc}
\end{figure}

Junto a los desafios tecnológicos y atmosféricos, podemos agregar que
los tres órdenes de magnitud del IR tienen en común el origen de la
emisión del Sol calmo, dominada por bremsstrahlung térmico
electrón--H$^+$ y, en menor medida, electrón--H$^0$. En esta atmósfera
en LTE la función fuente es la función de Planck que además puede ser
aproximada por Rayleigh-Jeans $S_\nu \simeq 2 k_B T\nu^2/c^2$ (con
$k_B$ la constante de Boltzmann, $T$ la temperatura de brillo y $c$ la
velocidad de la luz en el vacío). Esto significa que el flujo
observado es directamente proporcional a la temperatura de brillo, lo
que simplifica las interpretaciones \citep{Wedemeyeretal:2016}. Y en
los casos ópticamente delgados la temperatura de brillo corresponde a
la temperatura del plasma. Además, los modelos atmosféricos muestran
que el continuo de diferentes frecuencias se forma a diferentes
alturas lo que permite hacer diagnósticos a lo largo de la fotosfera /
cromosfera. En la Fig. \ref{fig:valc} vemos las diferentes alturas a
las que se forman las diferentes frecuencias. Estas alturas se
modifican durante las fulguraciones corriéndose a alturas mayores
\citep{Simoesetal:2017}. También durante las fulguraciones, se suma al
continuo submm la emisión girosincrotrónica producida
por electrones acelerados (o \textit{supratérmicos}) durante su
movimiento espiral en torno a las líneas magnéticas
\citep{PickVilmer:2008}.\\

\subsection{Líneas espectrales}

Que la cromosfera es una región inhomogenea queda comprobado por la
existencia de líneas espectrales moleculares muy intensas, como CO
$4.666\ \mu$m.  Los campos magnéticos pueden ser observados
por medio del efecto Zeeman en la línea de Fe~\textsc{i} $1.564\ \mu$m
y durante fulguraciones en la línea de He~\textsc{i} 1.083~$\mu$m,
aunque esta puede formarse hasta 2000~km de altura, ya en la región de
transición \citep{Avrettetal:1994}.  En la región Media del rango IR,
la línea de Mg~\textsc{i} $12.318\ \mu$m tiene la más alta
\textit{sensibilidad magnética} y permite la reconstrucción completa del
campo magnético cromosférico si se cuenta con un espectropolarímetro
que resuelva los 4 parámetros de Stokes \citep{Demingetal:1994}. \\

La región submm cuenta también con líneas espectrales de
H~\textsc{i} para transiciones atómicas de niveles altos $n \ge 19$, o
incluso del C~\textsc{iii}. El modelado de estas líneas no está
completamente resuelto, en parte por la complejidad de la cromosfera y
en parte por la carencia de observaciones que sienten 
restricciones \citep{Wedemeyeretal:2016, Clarketal:2000, Clarketal:2000b}.

\section{Observaciones}

La región del IR Próximo cuenta con una extensa literatura producto de
años de observación con telescopios cada vez más refinados. De hecho
en este momento se está realizando el comisionamiento del Telescopio
Solar Daniel K. Inouye (DKIST) localizado en el observatorio de
Haleakala (Hawaii, USA) a 3800 m sobre el nivel del mar. Este
telescopio tiene una apertura de 4~m y un rango especral que cubre
desde el azul hasta el IR Próximo, 0.38--5~$\mu$m, logrando
resoluciones espaciales de entre 0.02 y 0.24 segundos de arco. En su
foco cuenta (o contará) con una amplia oferta de detectores. Siendo
que las observaciones en el IR Próximo son más fotosféricas, no nos
referiremos aquí a esta rango.\\

\subsection{El Continuo del Sol calmo y activo}

\subsubsection{Sol Calmo}

\begin{figure}
\centerline{\includegraphics[width=0.3\textwidth]{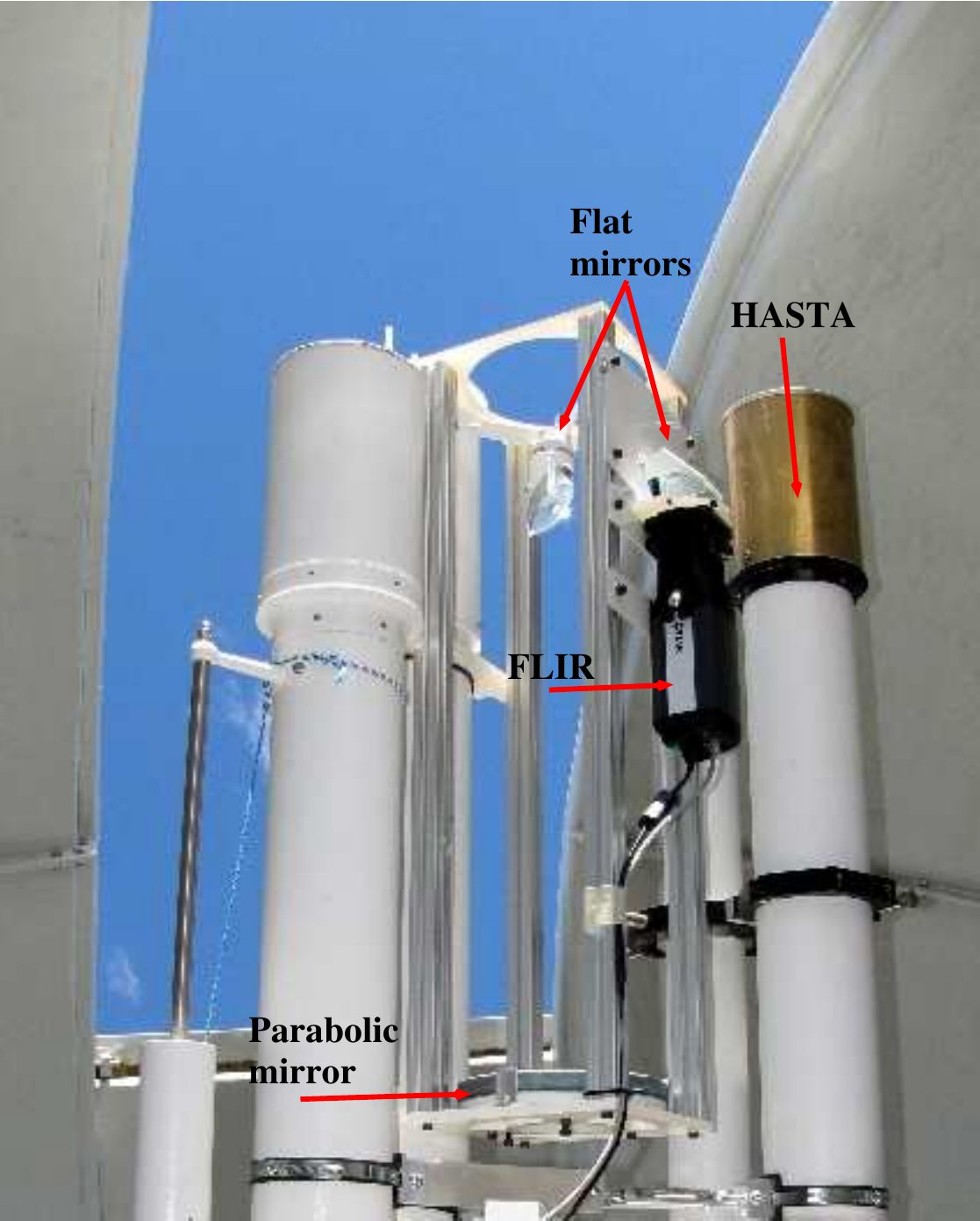}}
\caption{AR30T está formado por un telescopio newtoniano de 20 cm de
  apertura y una cámara FLIR. El sistema está adosado al telescopio
  HASTA.}
\label{fig:ar30t}
\end{figure}

Desde la década de 2000, el Centro de Rádio Astronomia e Astrofísica
Mackenzie (CRAAM) viene utilizando cámaras comerciales con detectores
no refigerados centrados en 10~$\mu$m y banda pasante de
aproximadamente 2~$\mu$m. Estos detectores, que se han mostrado
suficientemente sensibles para observaciones solares, fueron adosados
al foco de telescopios reflectores, en algunos casos usando también
celostatos
\citep{Marconetal:2008,Cassianoetal:2010,Kudakaetal:2015}. En este
momento el CRAAM posee dos configuraciones permanentemente montadas:
el SP30T en São Paulo \citep{Kudakaetal:2015,
  GimenezdeCastroetal:2018} y el AR30T en Argentina
\citep{Lopezetal:2022}. Ambas configuraciones comparten
características: resolución espacial del orden de 15 segundos de arco,
sensibilidad del orden de 1~K, resolución temporal de 1~s. SP30T usa
un celostato y un telescopio newtoniano de 15 cm de apertura para
crear un FOV $> 0.5^\circ$ en la cámara FLIR Termovision AM20 de $160
\times 120$ pixeles.  Por su parte AR30T tiene un espejo parabólico de
20 cm de diámetro, está adosado al telescopio HASTA y
usa una cámara FLIR Termovision SC645 de $640 \times 480$ pixeles con
un FOV $\approx 20$ minutos de arco (Fig. \ref{fig:ar30t}).\\

\begin{figure}
\centerline{\includegraphics[width=0.5\textwidth]{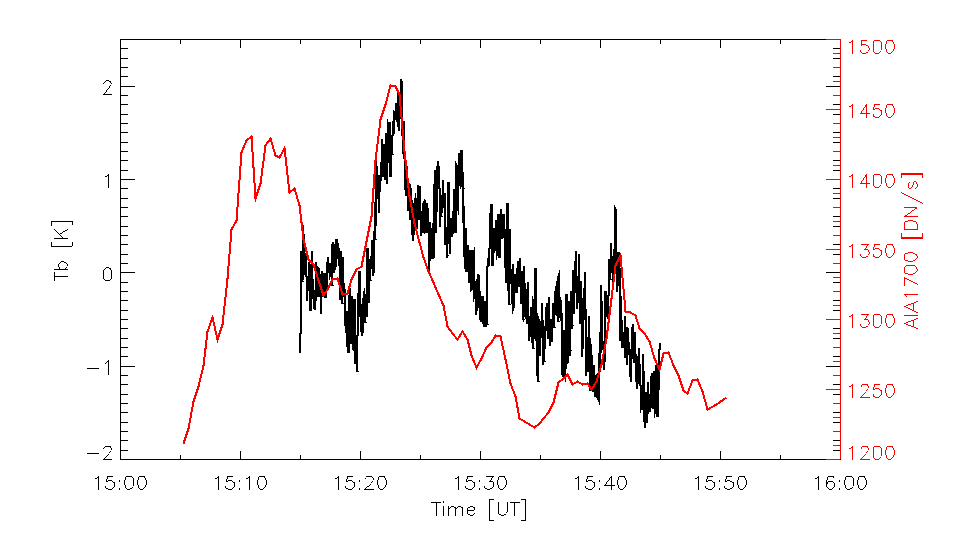}}
\caption{Intensidad luminosa registrada por AR30T en 10~$\mu$m durante
  SOL2022-02-28T15:20 (negro) superpuesta con la intensidad en
  1700~\AA\ (rojo). (Gentileza de F. López, Univ. de Mendoza/CONICET,
  trabajo en preparación).}
\label{fig:sol2022-02-28}
\end{figure}

Observaciones en 10~$\mu$m muestran una gran correlación espacial con
líneas espectrales \citep{Marconetal:2008}. Durante fulguraciones la
emisión es compatible con bremsstrahlung
térmico. \cite{Trottetetal:2015} usan el modelo atmosférico de
\cite{Machadoetal:1980} y concluyen que la emisión proviene de una
altura de $\approx 1000$~km sobre la fotosfera y originada en el
calentamiento del plasma provocado por la precipitación de las
partículas aceleradas. \cite{Pennetal:2016} observaron el evento
SOL204-09-24T17:50 en 5.2 y 8.2~$\mu$m usando un detector refrigerado
de tipo QWIP en el foco del telescopio MacMath-Pierce consiguiendo una
resolución de 3 segundos de arco. Estas observaciones simultáneas en
longitudes de onda diferentes confirman la tesis de que la emisión es
bremsstrahlung térmico ópticamente delgado. En general, durante las
fulguraciones se observa una correlación espacio-temporal con la
emisión en luz blanca \citep{Kaufmannetal:2013} y con el continuo en
1700 y 1600~\AA\ \citep{Lopezetal:2022}. Esta última característica se
revela incluso en eventos muy débiles como es el caso de
SOL2022-02-28T15:20 (Fig. \ref{fig:sol2022-02-28}), clasificado de
B2 por su emisión en rayos-x blandos. \\

El IR Lejano ha sido muy poco explorado, podemos apenas encontrar dos
intentos esporádicos. Su mayor dificultad reside en la necesidad de
observaciones por fuera de la troposfera terrestre. Las primeras
observaciones se remontan a 1987 utilizando el Kuiper Airbone
Observatory de NASA \citep{Lindseyetal:1990}, un avión que volaba a
12.800 m sobre el nivel del mar con un telescopio de 1 m de apertura
al que los autores adicionaron detectores para las longitudes de onda
de 50, 100 y 200~$\mu$m. Se reportaron variaciones de varios K en la
temperatura del Sol calmo, asociadas con oscilaciones de 5 minutos
observadas en la línea D1 del sodio (5894~\AA). Casi 30 años más
tarde, \cite{Kaufmannetal:2016} construyeron un telescopio, el
Solar-T, que voló adosado al globo estratosférico Gamma-Ray Imager /
Polarimeter for Solar Flares (GRIPS), con detectores bolométricos
basados en células de Golay y filtros pasa banda para observar el Sol
en las longitudes de onda de 43 y 100~$\mu$m.\\

\begin{figure}
\centerline{\includegraphics[width=0.3\textwidth]{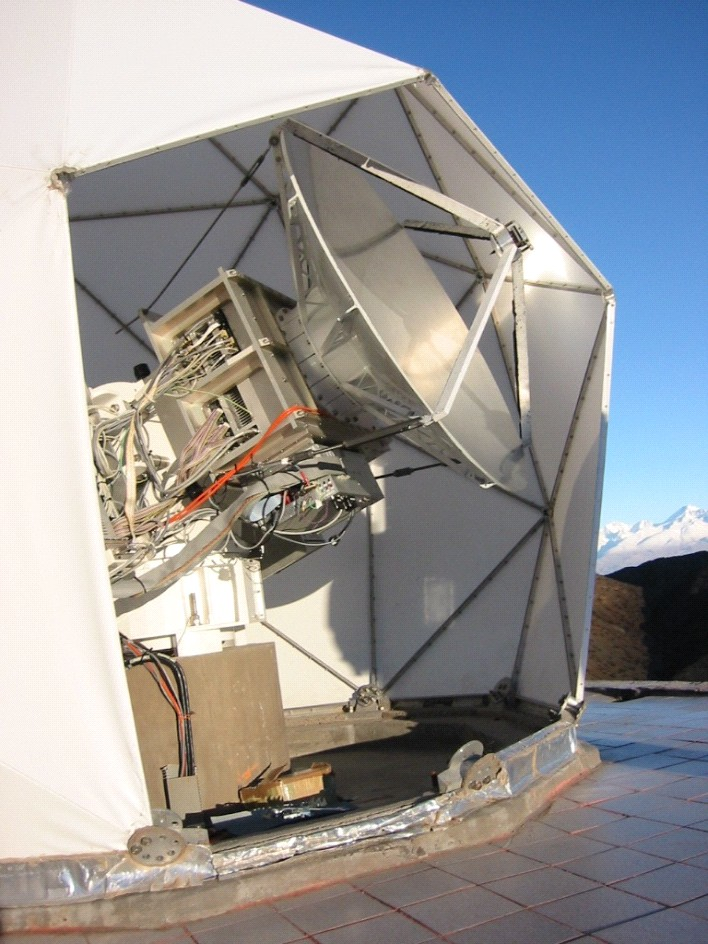}}
\caption{El SST con el radomo abierto para tareas de mantenimiento.}
\label{fig:sst}
\end{figure}

El submm fue ocasionalmente observado con el radiotelescopio de 15~m
de apertura James Clerk Maxwell Telescope \citep[JCMT,
][]{LindseyKopp:1995,Lindseyetal:1995} en 350~$\mu$m (857~GHz),
850~$\mu$m (353~GHz) y 1200~$\mu$m (250~GHz), revelando la existencia
de abrillantamiento al limbo en todas las frecuencias aunque con
intensidades diferentes. También mostraron la estructura de las
manchas: la penumbra tiene temperaturas semejantes a las plages
vecinas, mientras que la umbra es más de 1000~K más fría. \\

El Telescopio Solar Submilimétrico \citep[SST][]{Kaufmannetal:2008},
instalado en el CASLEO a 2500~m sobre el nivel del mar, es el primero
en observar rutinariamente el sol en 740~$\mu$m (405~GHz) y
1415~$\mu$m (212~GHz). Este radiotelescopio Cassegrain de 1.5 m de
diámetro (Fig. \ref{fig:sst}), cuenta con un sistema multi haz que le
permite localizar el centroide de emisión de una fuente puntual con
una resolución temporal de hasta 5~ms
\citep{GimenezdeCastroetal:1999}. Los seis haces que componen el
arreglo focal tienen tamaños angulares nominales de 2 y 4 minutos de
arco para 740 y 1415~$\mu$m, respectivamente y una sensibilidad del
orden de 10~K. En un análisis provisorio que incluyó apenas una
veintena de mapas, \cite{Silvaetal:2005} encontraron que la emisión de
las regiones activas es compatible con bremsstrahlung térmico en el
régimen ópticamente grueso. Esta emisión debe provenir mayormente de
las \textit{plages} vecinas y la penumbra ya que por la resolución
espacial del SST la emisión observada resulta de la convolución de
\textit{plage}, penumbra y umbra. Este resultado fue confirmado
posteriormente por \cite{ValleSilvaetal:2021} adicionando imágenes
de baja resolución espacial del Atacama Large Millimeter Array
(ALMA) en 3000~$\mu$m y por \cite{GimenezdeCastroetal:2020b}
aprovechando los 20 años de observaciones diarias del SST; además,
estos últimos autores mostraron que el exceso de temperatura de brillo
de las regiones activas está correlacionado positivamente con el ciclo
solar. Más recientemente \cite{Menezesetal:2021, Menezesetal:2022}
midieron con gran precisión el radio solar usando el SST y mostraron
su variación a lo largo del ciclo.\\

\begin{figure}
\centerline{\includegraphics[width=0.4\textwidth]{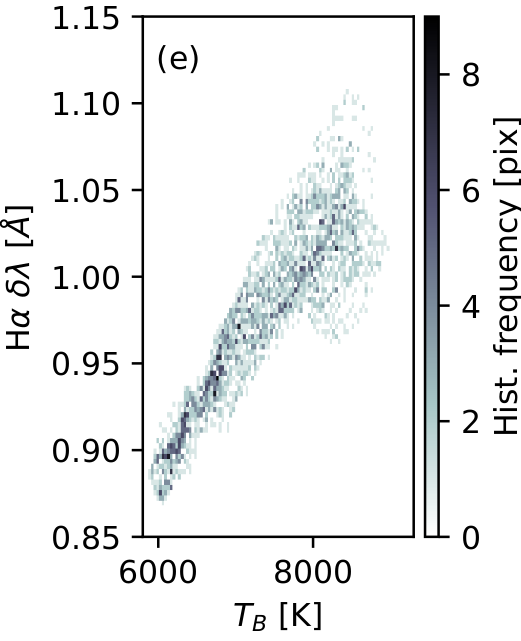}}
\caption{Distribución de probabilidad conjunta de la temperatura de
  brillo en 3~mm y el ancho de la línea H$\alpha$. \citep[Reproducido
    con permiso de ][]{Tarretal:2023}}
\label{fig:tarr}
\end{figure}

Usando observaciones de ALMA, se han corroborado
muchas de las observaciones realizadas con el JCMT
\citep{Whiteetal:2017, Alissandrakisetal:2017}. En modo
interferómetro, ALMA alcanza resoluciones espaciales $< 0.01$~segundos
de arco lo que le permite estudiar detalles hasta ahora desconocidos
en este rango. \cite{Loukitchevaetal:2017} consiguen separar la
emisión de las distintas partes de una mancha, mostrando que hay un
comportamiento diferente en 1.3~mm respecto de 3~mm. La correlación
con emisión espectral, particulamente con la línea de H$\alpha$, es
analizada por \cite{Molnaretal:2019} y posteriormente confirmada por
\cite{Kobelskietal:2022} y \cite{Tarretal:2023}. La mejor correlación
espacio temporal se da entre observaciones en 3~mm y el ancho de la
línea H$\alpha$ y hay también proporcionalidad entre el ancho de
H$\alpha$ y la temperatura de brillo del continuo de 3~mm (Fig.
\ref{fig:tarr}). Estas correlaciones pueden deberse a la sensibilidad
de H$\alpha$ al ensanchamiento Doppler térmico \citep{Cauzzietal:2009}
aunque de ser así, cuestionan \cite{Molnaretal:2019}, implicaría
temperaturas por encima de los 50000~K y una cromosfera completamente
ionizada.\cite{Patsourakosetal:2020} observan las oscilaciones de 3--5
minutos sobre el Sol calmo, mientras que \cite{Tarretal:2023} muestran
evidencia marginal de estas oscilaciones sobre las manchas.\\

\subsubsection{Fulguraciones}

El mayor impacto del SST han sido sus observaciones de fulguraciones,
mostrando una íntima correlación con la emisión en rayos-$\gamma$
\citep{Kaufmannetal:2002} y rayos-X duros
\citep{GimenezdeCastroetal:2009}, mientras que el carácter
discretizado de la emisión fue analizado por \cite{Raulinetal:2003} y
\cite{Kaufmannetal:2009}. Usando una configuración parecida al SST, el
radiotelescopio Kölner Observatorium für Submillimeter Astronomie
(KOSMA) de 3 m de apertura tuvo por un breve tiempo un sistema
multi-beam que le permitió determinar, también, el tamaño de la fuente
emisora y su variación durante la fulguración
SOL2003-10-28. \cite{Luthietal:2004b} muestran que el tamaño de la
fuente emisora llega a ser menor que la resolución teórica del
instrumento de unos pocos segundos de arco. Las observaciones de KOSMA
de este evento fueron analizadas por \cite{Trottetetal:2008} junto con
imágenes de rayos-$\gamma$ y rayos-x, mostrando que el submm coincide
espacialmente con los primeros, resultado que refuerza la tesis de que
la emisión submm se origina en partículas de muy alta energía.\\

\begin{figure}
  \centerline{\includegraphics[width=0.2\textwidth]{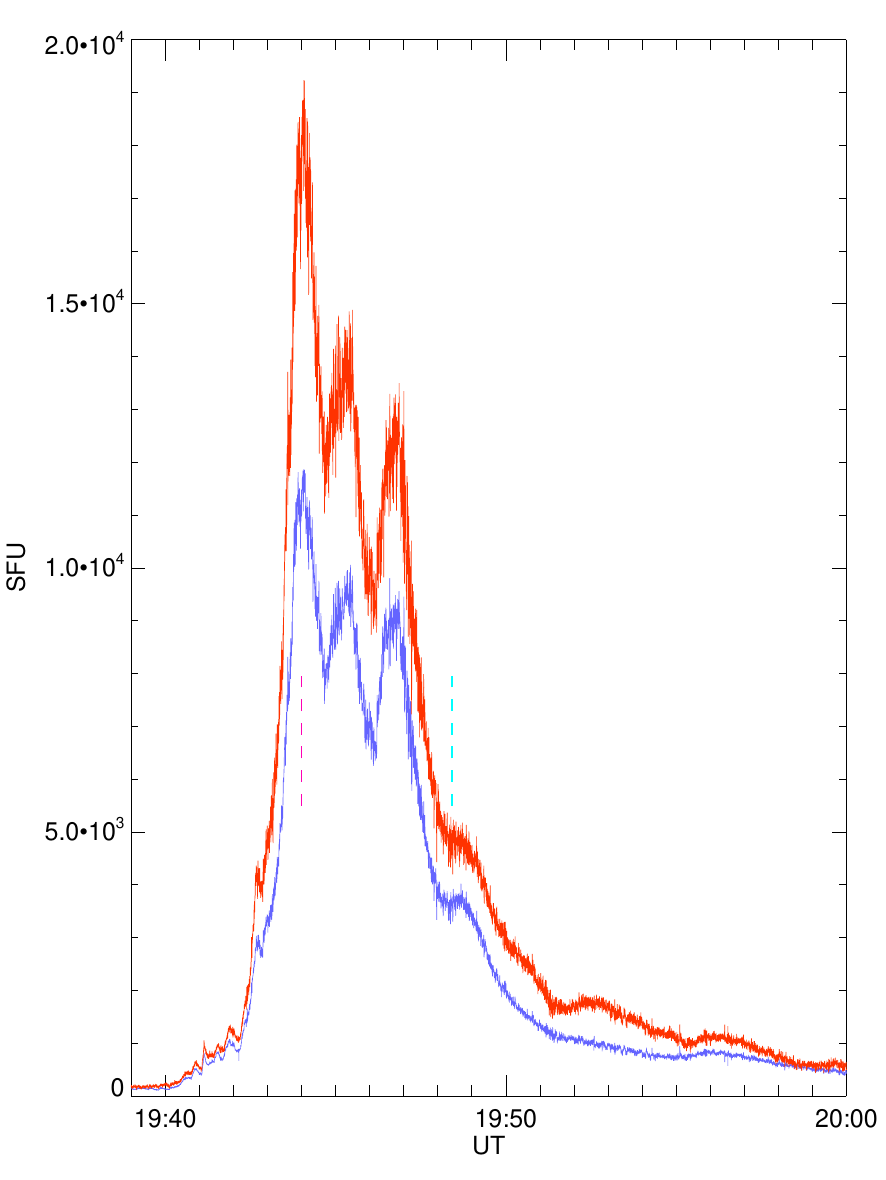}
  \includegraphics[width=0.2\textwidth]{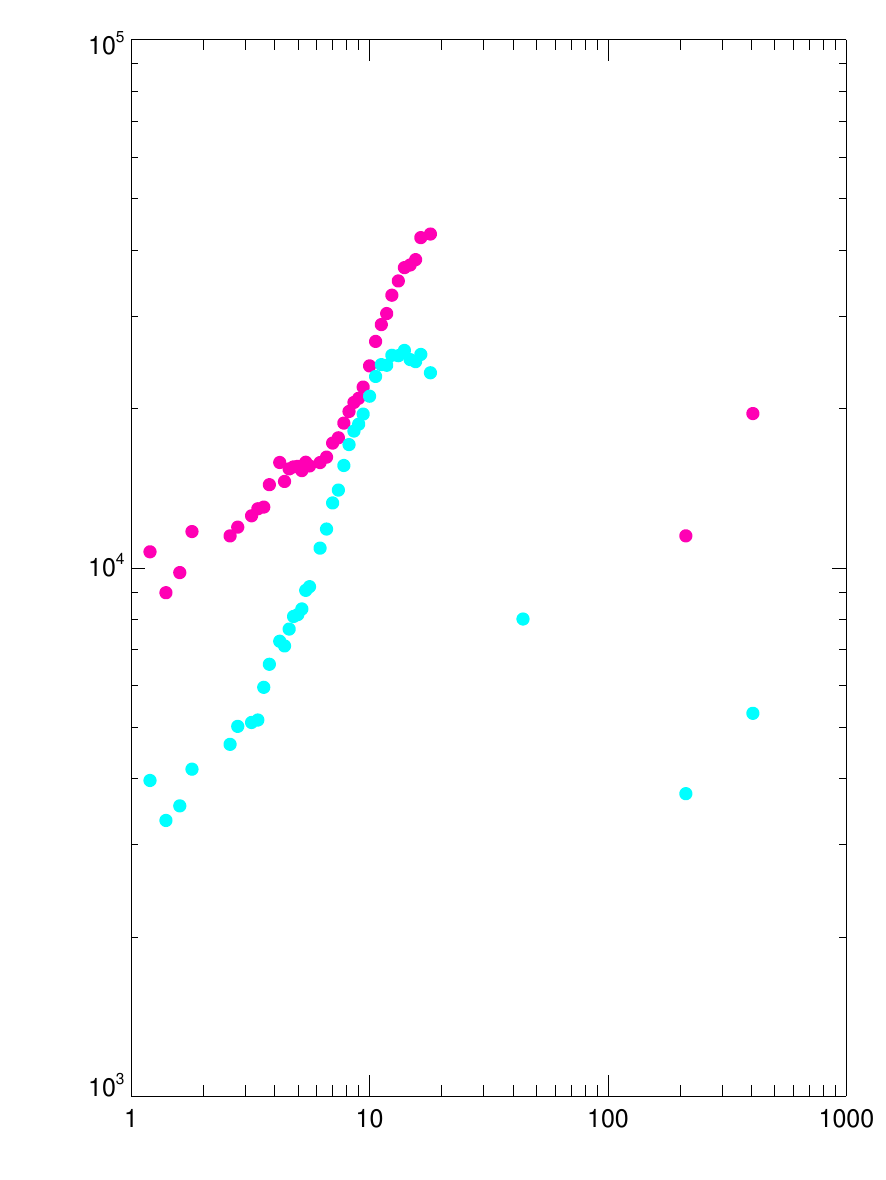}}
\caption{Izquierda: variación de la densidad de flujo durante el
  evento SOL2003-11-04T19:45, rojo, 405~GHz, azul, 212 GHz. Derecha:
  dos espectros tomados en instantes diferentes mostrando la
  componente por encima de 200~GHz diferenciada de las frecuencias más
  bajas.}
\label{fig:sol2003-11-04}
\end{figure}

La existencia de una componente espectral submm continua diferente a
la observada en microondas observada en varios eventos es aún materia
de investigación \cite[Fig. \ref{fig:sol2003-11-04},
][]{Kaufmannetal:2004}. El origen es especulado en
\cite{Kruckeretal:2013} e incluye, además de bremsstrahlung térmico y
síncrotron de electrones ultrarelativísticos, síncrotron de positrones
producidos durante reacciones nucleares y otras causas más exóticas.
A pesar de más de 20 años de la primera detección de un ``evento
  THz'', la instrumentación todavía no permite llegar a una conclusión
firme porque la respuesta a la incógnita sólo se logrará por medio de
observaciones simultáneas en frecuencias mayores incluyendo
polarización e imágenes.\\

\subsection{Líneas espectrales}

Observaciones de líneas espectrales desde el IR Medio hasta el submm
son muy escasas aunque su potencial es muy
grande. \cite{Hewagamaetal:1993} usaron un espectropolarímetro de alta
resolución espectral y un detector IR de $128 \times 128$ pixeles para
observar el efecto Zeeman de la línea de Mg~\textsc{i}
12.318~$\mu$m. Aplicando un modelo NLTE para calcular la tranferencia
radiativa de los parámetros de Stokes, \textit{I, Q, U} y \textit{V},
mostraron que el perfil de la línea no es afectado por la saturación y
por lo tanto se puede obtener el campo magnético vectorial con alta
confiabilidad incluso para intensidades muy grandes. \\

También se han estudiado muy poco las líneas espectrales del submm con
la única excepción de los trabajos de
\cite{Clarketal:2000,Clarketal:2000b} quienes aprovecharon las pocas
oportunidades en que el JCMT observó el Sol en la década de 1980, en
337~$\mu$m (888~GHz) y 453~$\mu$m (662~GHz). En el primer artículo,
estos autores muestran claramente la línea de H~\textsc{i} en
absorción con un efecto de abrillantamiento al limbo y mezclada con
una posible línea de Mg~\textsc{i}. En el segundo artículo, la línea
es sólo observada próxima al limbo. Estudios espectrales en el submm
permitirán establecer con mayor precisión la escala de temperatura en
la cromosfera y región de transición, mientras que los perfiles pueden
ser usados para estudiar la turbulencia, el flujo y la densidad del
plasma \citep{Wedemeyeretal:2016}.

\section{Instrumentación: presente y futuro}

A lo largo de este artículo ya hemos mencionado la mayoría de los
instrumentos en funcionamiento para los rangos del IR Medio al submm,
casi todos operados por el CRAAM en cooperación con instituciones
argentinas como CASLEO y OAFA: las cámaras AR30T (OAFA) y SP30T (São
Paulo), y el SST (CASLEO) al que podemos agregar como instrumento
auxiliar el polarímetro Polarization Emission of the Millimeter
Activity at the Sun (POEMAS) \citep[en el CASLEO, ][]{Valioetal:2013}
que observa en 3.3 y 6.6~mm (90 y 45~GHz, respectivamente). No hay
otros instrumentos dedicados al Sol en esta banda de
frecuencias. Eventuales observaciones han sido realizadas con el JCMT,
como hemos dado cuenta. Ya en el borde del submm (3.2 y 2.2~mm) con el
telescopio ruso RT-7.5 \citep{Tsapetal:2018}, y en 2.6 y 3.5~mm con la
gran antena de 45~m de apertura de Nobeyama
\citep{IwaiShimojo:2015}. El interferómetro ALMA tiene como uno de sus
\textit{use cases} observaciones solares. Su uso, sin embargo, ha
mostrado que es adecuado principalmente para analizar las estructuras
atmosféricas del Sol calmo y quiescente con gran detalle. La
metodología de observación, que precisa de calibraciones con duración
de algunos minutos a cada 10 minutos aproximadamente, lo hace poco
útil para la detección de las oscilaciones fotosféricas de 3--5
minutos \citep[ver las conclusiones de ][por ejemplo]{Tarretal:2023},
mucho menos para fulguraciones cuyo carácter espacio-temporal es
aleatorio. Se suma a estos inconvenientes, la baja oferta de tiempo
para observaciones solares (en 2022 sólo un proyecto fue aprobado),
limitación de recursos (polarización lineal unicamente, pocas bandas
espectrales) y la imposibilidad de observaciones simultáneas en varias
frecuencias del continuo, lo que limita la obtención de espectros
confiables \citep[ver, por ejemplo, ][ quienes analizan una erupción
  de plasma]{Rodgeretal:2019}. Curiosamente, ALMA ha inaugurado,
\textit{serendípicamente}, la observación submm de fulguraciones
estelares \citep{MacGregoretal:2018,MacGregoretal:2020}. \\

\begin{figure}
  \centerline{\includegraphics[width=0.4\textwidth]{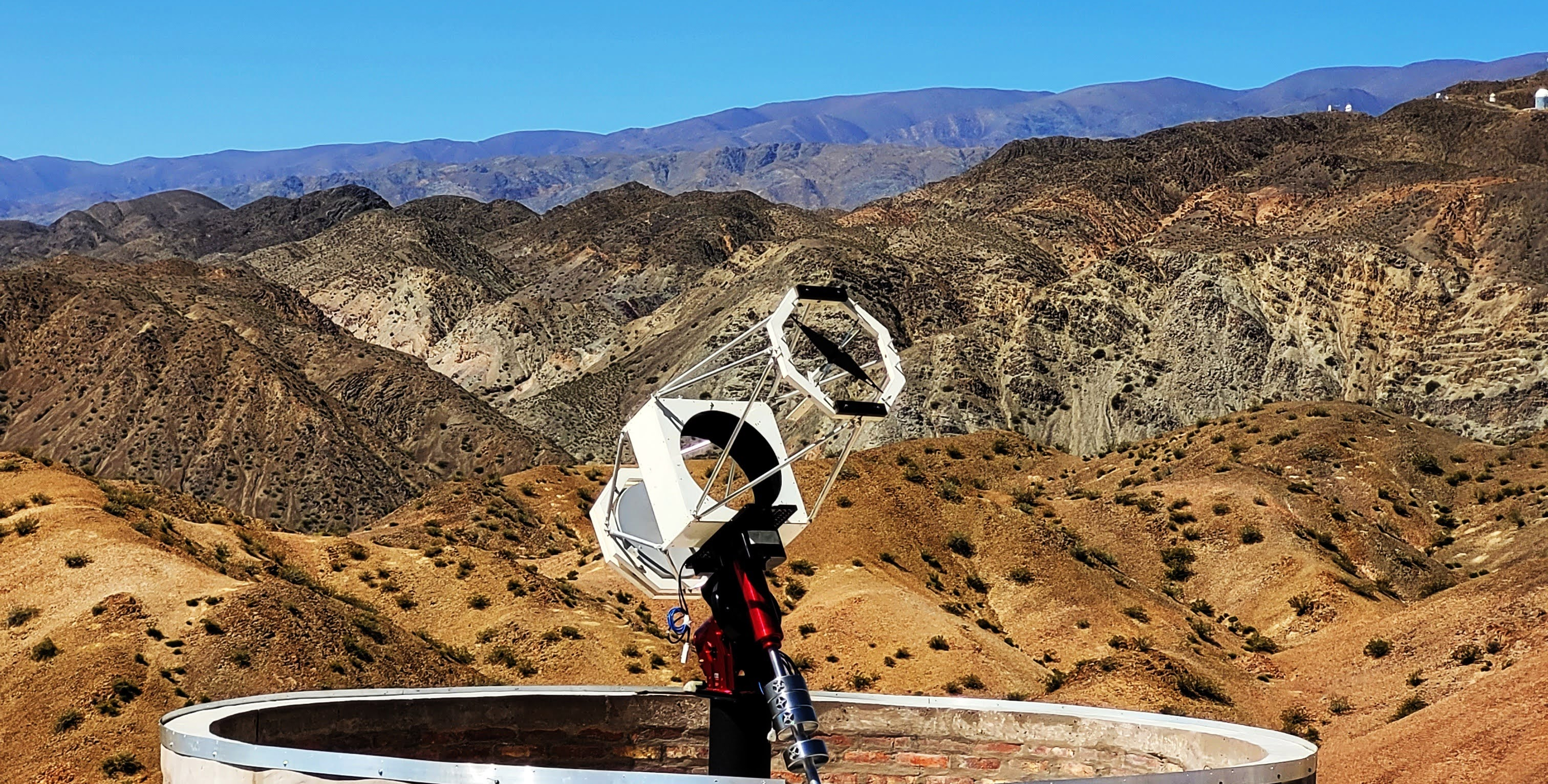}}
  \centerline{\includegraphics[width=0.5\textwidth]{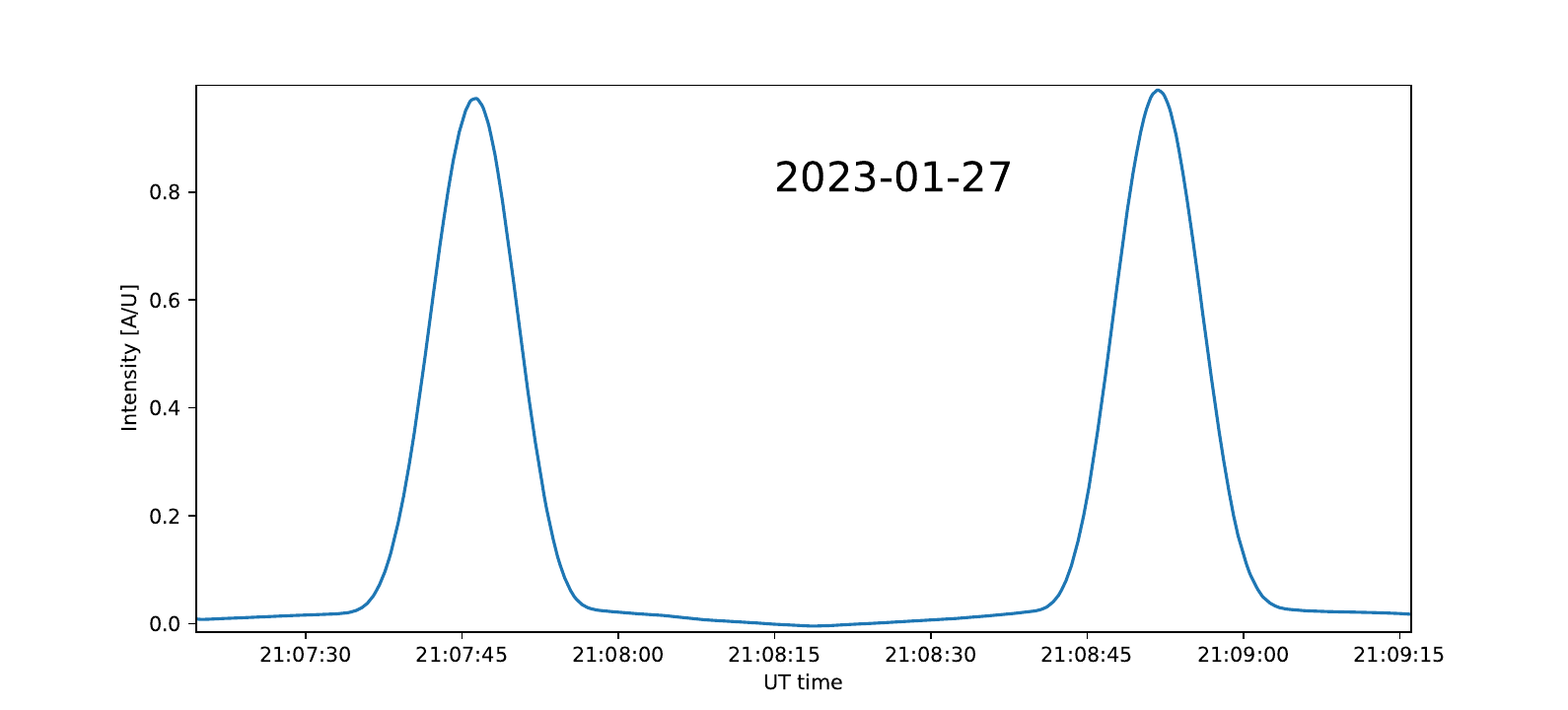}}
  \caption{Arriba: HATS durante la etapa de integración y tests en
    enero de 2023. Abajo: Perfiles solares obtenidos por barridas en
    las direcciones de ascención recta (izquierda) y declinación
    (derecha). Primeras observaciones no-calibradas de HATS.}
\label{fig:hats}
\end{figure}

En enero de 2023 el \textit{High Altitude THz Solar photometer}
\citep[HATS, ][]{GimenezdeCastroetal:2020} fue integrado en su lugar
de instalación la Estación Carlos U. Cesco del OAFA (Fig.
\ref{fig:hats}), y pasó satisfactoriamente los tests de
funcionamiento. HATS cuenta con un detector centrado en la longitud de
onda de 20~$\mu$m (15~THz) en el foco primario de un telecopio de
montura ecuatorial con 45~cm de apertura y un haz de de gran tamaño
para observar el Sol entero. La figura \ref{fig:hats} presenta las
primeras barrida no calibradas sobre el Sol, que para
todos los efectos representan el perfil del haz (el Sol es mucho menor
que el haz) y demuestran la calidad de la óptica que no genera
deformaciones en la imagen. En este momento el telescopio entró en la
fase de comisionamiento mientras realiza las primeras observaciones
solares en esta frecuencia todavía inexplorada. \\

El proyecto binacional argentino-brasilero \textit{Large Latin
  American Millimeter Array} \citep[LLAMA,
][]{Lepineetal:2020,Arnaletal:2017} tendrá capacidad de realizar
observaciones solares en diferentes bandas de frecuencias
simultáneamente, una característica especialmente diseñada para
capturar eventos transientes solares, convirtiéndolo en un instrumento
único en el submm. Su construcción, largamente demorada, ha ganado
fuerte impulso y se espera que para 2024/2025 comience la etapa de
comisionamiento y primeras observaciones científicas.\\

Por último, el SST, único telescopio solar submm en el mundo, debe
pasar por una reforma integral que lo tornará más sensible cambiando
las frecuencias de observacion a las  más adecuadas para el sitio y
aumentando el tamaño del reflector primario, contará con polarización
circular y un espectrómetro. El SST \textit{next generation} se
encuentra ahora en la etapa de diseño, esperamos contar con
financiamiento en los próximos años para comenzar su construccion.\\

\section{A modo de conclusión}

El rango espectral que abarca desde el IR Medio al submm, es decir de
5 a 1000~$\mu$m, ha sido muy poco explorado. Aumentar la cobertura
espectral, contar con medidas de polarización circular y de líneas
espectrales con instrumentos dedicados exclusivamente al Sol mejorará
los diagnósticos y nuestra comprensión de los procesos
cromosféricos. Argentina dispone en la actualidad de instrumentos
únicos a los que se le sumarán en el futuro próximo otros más, con los
que puede realizar aportes significativos en el área. AR30T, por
ejemplo, ha demostrado capacidad de observar eventos muy débiles, lo
que abre perspectivas de colaboración con otros instrumentos (por
ejemplo, análisis conjuntos con el Spectrometer Telescope for Imaging
X-rays (STIX), detector de rayos-x a bordo del Solar Orbiter) y contar
con una estadística mayor de casos. Por otro lado, la excelente
correlación entre 10~$\mu$m, luz blanca y el continuo UV, puede
convertir al IR Medio en un proxy de las últimas dos usando
instrumentos en tierra. \\

Los rectángulos coloreados de la Fig. \ref{fig:valc} muestran el rango
de alturas aproximado de las observaciones de los diferentes
instrumentos para un modelo de Sol calmo. Como se puede ver, la región
del IR Lejano que permitiría observar la tarnsición entre fotosfera y
cromosfera, sigue estando despoblada de instrumentos. La comunidad
solar internacional debería hacer un esfuerzo por colocar un
instrumento para esta banda a bordo de un satélite o estación
espacial. Infelizmente, varios intentos anteriores, que vienen desde
la década de 1980\footnote{Una versión preliminar del Satélite de
  Aplicaciones Científicas B, (SAC-B), de la Comisión Nacional de
  Aplicaciones Espaciales, contemplaba un telescopio solar para el
  IR.} no consiguieron financiamiento. Creemos que esta es la hora de
completar el espectro solar con observaciones que nos han de llenar de
nuevas informaciones y preguntas.

\begin{acknowledgement}
  Agradesco a la Asociación Argentina de Astronomía, en particular al
  Comité Científico de la 64ta reunión, por la gentil invitación a dar
  una charla invitada. Agradesco además a la FAPESP, CAPES,
  Mackpesquisa y CNPq de Brasil y al CONICET y la Universidad Nacional
  de San Juan de Argentina por el apoyo financiero y humano recibido
  para realizar las investigaciones cuyos resultados se presentan
  aquí. Especial agradecimiento a Lucas Tarr (NSO, USA) que cedió la
  figura \ref{fig:tarr}. Por último, agradesco a mis colegas del CRAAM
  en Brasil y del IAFE en Argentina que no están incluidos en la lista
  de autores, porque sin su colaboración de más de 20 años, este
  artículo no habría jamás existido.
\end{acknowledgement}

\bibliographystyle{baaa}
\small
\bibliography{referencias}

\end{document}